\documentclass[11pt,dvips]{article}
\usepackage{epsfig,times} 
\usepackage{epsfig}
\usepackage{picinpar}
\usepackage{wrapfig}
\usepackage{floatflt}
\makeatletter
\renewcommand{\section}{\@startsection{section}{1}{0in}
	{0.4\baselineskip}{0.1\baselineskip}{\Large\bf}}
\renewcommand{\subsection}{\@startsection{subsection}{2}{0in}
	{0.25\baselineskip}{-\baselineskip}{\large\bf}}
\renewcommand{\subsubsection}{\@startsection{subsubsection}{3}{0in}
	{0.1\baselineskip}{-\baselineskip}{\normalsize\bf}}
\makeatother
\begin{document}
%
%  Session and Paper Code:
\thispagestyle{myheadings}
\markright{HE.1.3.13}
\begin{center}
{\LARGE \bf Inelastic proton-air cross section in U.H.E.}
\end{center}

%  Author List:
\begin{center}
{\bf J.R. Fleitas, J. Bellandi}\\
{\it Instituto de F\'{\i}sica, Universidade Estadual de Campinas, Campinas, SP 13083-970, Brazil }
\end{center}

%  Abstract:
\begin{center}
{\large \bf Abstract\\}
\end{center}
\vspace{-0.5ex}
We calculate the inelastic proton-air cross section $\sigma_{in}^{p-air}$ 
$(mb)$ by means of the Glauber model. As input, we use acceletator experimental data on the total proton-proton cross section. We present also a paramentrization with the energy for this cross section and compare with cosmic ray experimental data and with Monte Carlo simulation.
%

%  Leave this line skip in place:
\vspace{1ex}

%
%  Manuscript text:
%

In this paper we determine the inelastic  $p-air$ and ${\bar p}-air$ cross section in
the Glauber framework (Glauber 1959; Glauber et al. 1970) using experimental data on the total $
pp$ and ${\bar p}p$ cross section (Caso et al. 1998) and compare with experimental data cross
sections from Akeno Collab. (Honda et al. 1993), Fly's Eyes Collab. (Baltrusaitis et al. 1985) and EASTOP Collab. (Aglietta et al. 1998). In the Glauber model  the relationship between the
inelastic hadron-air cross section. The $\sigma _{in}^{h-air}$ $(h\equiv p,{\bar p})$ and total $hp$ cross section $\sigma _{tot}^{hp}$ , is given by 
\begin{equation}
\sigma _{in}^{h-air}=\int d^{2}b\left[ 1-\exp [-\sigma _{tot}^{hp}AT(\mathbf{%
b})]\right]   \label{1}
\end{equation}
where {\bf b} is the impact parameter, $T(b)$ is the nuclear thikness, 
\begin{equation}
T(b)=\int dz\rho (b,z)  \label{2}
\end{equation}
given in terms of the nuclear distribution $\rho (b,z).$

In order to calculate the nuclear thickness $T(b)$ we use here the
Woods-Saxon model (Woods \& Saxon 1954; Barrett \& Jackson 1977) for the nuclear distribution which is given by 
\begin{equation}
\rho (r)=\rho _{o}\left[ 1+\exp [\frac{r-r_{o}}{a_{o}}\right] ^{-1}(1+\omega 
\frac{r^{2}}{r_{o}^{2}})  \label{3}
\end{equation}
where the factor $(1+\omega \frac{r^{2}}{r_{o}^{2}})$ correspond to the
Fermi parabolic distribution correction. The parameter $\rho _{o}$ is a
normalization factor derived by means of 
\begin{equation}
\int d^{3}r\rho (r)=1  \label{4}
\end{equation}
The parameters $r_{o},a_{o}$ and $\omega $ can be derived from experimental
data (Barrett \& Jackson 1977) . Fitting the experimental data we have $r_{o}=0.976A^{1/3}$ fm , $%
a_{o}=0.546$ fm , and for the parameter $\omega $ we have 
\begin{eqnarray}
\begin{array}{c}
\omega =-0.25839,\quad{ if A}\leq 40 \\ 
\omega =0\quad{ if A}>40
\end{array}
\end{eqnarray}

The total proton-proton and antiproton-proton cross sections are well known
at low energies, but the highest energy data is undefined both from
accelerator and cosmic ray point of views. The last accelerator data came
from E710 (Abe et al. 1992) and CDF Collab. (Amos et al. 1994) and reported  discrepant measurements
of total anti-proton-proton cross section obtained at the Tevatron Collider (%
$\sqrt{s}=1.8$ TeV). From cosmic ray measurements, the last data from wich
one can derive nucleon-nucleon total cross section were reported by the
Akeno Collab. (Honda et al. 1993) and EASTOP Collab. (Agliettan et al. 1998).

We calculate here the $\sigma _{in}^{h-air}(mb)$ using equation (\ref{1}), and as input for $\sigma _{tot}^{h-p}(mb)$ we use accellerator experimental data
(Caso et al. 1998). In Figure 1 we show the calculated $\sigma _{in}^{h-air}(mb)$ as function
of $p_{lab}.$ We also show in this figure the following fit for the inelastic $%
h-A$ cross section $(h\equiv p$ and ${\bar p})$%
\begin{eqnarray}
\sigma _{in}^{h-A}=\sigma _{o}(p_{lab})A^{\alpha(p_{lab}) }  
\label{5}
\end{eqnarray}
where 
\begin{eqnarray}
\sigma _{o}(p_{lab})=a_{1}p_{lab}^{\epsilon }+a_{2}p_{lab}^{-\eta }
\label{6}
\end{eqnarray}
and 
\begin{eqnarray}
\alpha (p_{lab})=b_{1}(1+\frac{1}{p_{lab}})+b_{2}p_{lab}^{\xi }  
\label{7}
\end{eqnarray}
with $A=14.5.$ The values of the constants are given in Table 1. This
parametrization, at the Tevatron energy, goes between the values of $\sigma
_{in}^{h-A}$ as calculated using the values of $\sigma _{tot}^{{\bar p}p}$
from E710 (Amos et al. 1992)  and CDF (Abe et al. 1994) Collaborations, respectively.

\begin{center}
\begin{tabular}{|l||llll|l|} \hline  
reaction & $a_1 $ & $a_2$ & $\epsilon$ & $ \eta $ &  $\chi^2 $ \\
\hline \hline
$pp$ & $20.08\pm2.11$ & $27.51\pm1.28$ & $0.0852\pm0.0029$ & $-0.2045\pm0.0092$ & $3.21$  \\ 
${\bar p}p$ & $39.34\pm3.40$ & $27.51\pm1.28$ & $0.0852\pm0.0029$ & $-0.2045\pm0.0092$ & $3.21$ \\ 
\hline
\end{tabular}
\end{center}

\begin{center}
\begin{tabular}{|l||lll|l|} \hline  
reaction & $b_1$ & $b_2$ & $\xi $& $\chi^2 $ \\
\hline \hline
$pp$ & $-0.19357\pm0.01945$ & $0.91999\pm0.02093$ & $-0.00534\pm0.00022$ & $2.08\times 10^{-3}$ \\ 
${\bar p}p$ & $-0.19357\pm0.01945$ & $0.91999\pm0.02093$ & $-0.00934\pm0.00097$ & $2.08\times 10^{-3}$ \\ 
\hline
\end{tabular}
\end{center}

In Figure 2 we compare our calculated $\sigma_{in}^{p-air}$ $(mb)$ with results from Monte Carlo simulation in CORSIKA Code (Knapp et al 1996). Our calculation is in agreement with results from the hadron dual parton model (HDPM). In this figure we also show the values of $\sigma _{in}^{p-air}$ $(mb)$ as presented by the Akeno Collab. and the values as derived by Bellandi et al (Bellandi et al 1997) assuming single-diffractive and non diffractive
contributions to the hadronic flux in the atmosphere.

We would like to thank the Brazilian governmental agencies CNPq and CAPES for financial support.

\begin{figure}[htpb]
\vspace{-8cm}
\begin{center}
{\mbox{\epsfig{file=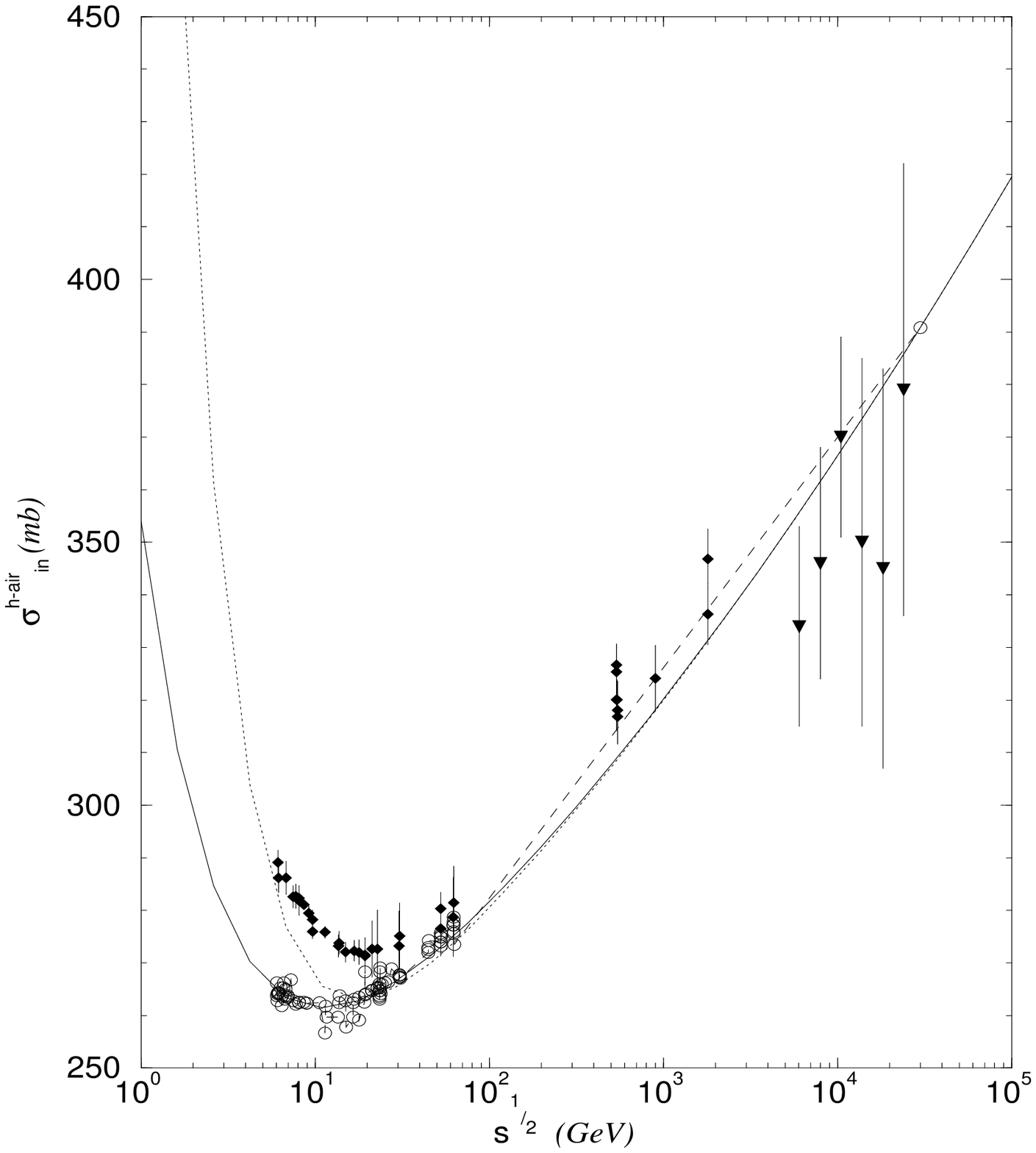,width=2.1in}}}
\end{center}
\caption{Open cicle is $p-air$ inelastic cross section. Close diamond is ${\bar p}-air$ inelastic cross section. Solid line is our paramentrization to $\sigma _{in}^{p-air}$ $(mb)$. Dot line is our parametrization to $\sigma _{in}^{{\bar p}-air}$ $(mb)$. Down triangle is Akeno data with Bellandi correction (Bellandi et al. 1997). Open diamond is EASTOP (Aglietta et al. 1998). Star is Fly's Eyes data (Baltrusaitis et al. 1985).  }

\begin{center}
{\mbox{\epsfig{file=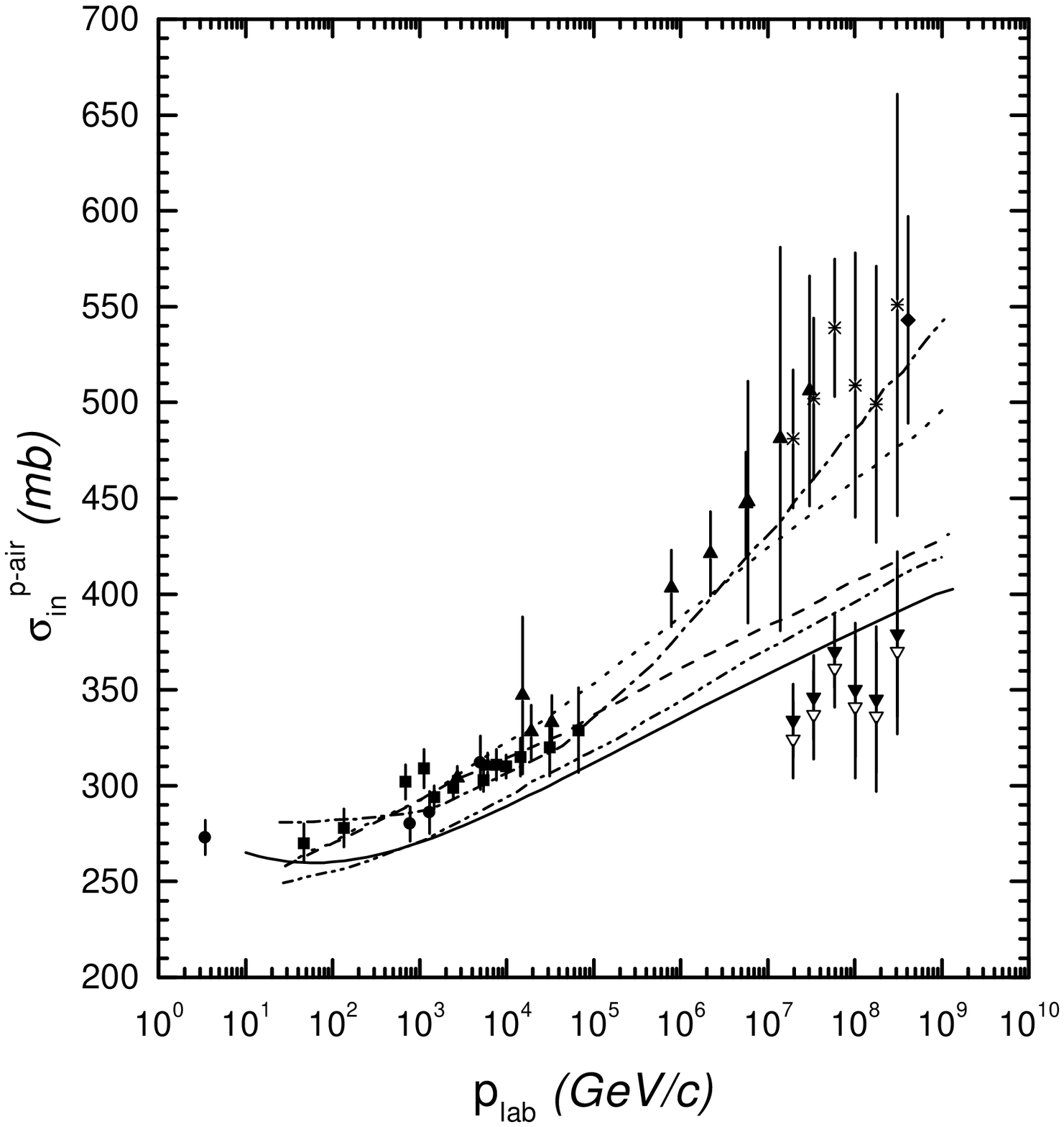,width=3.6in}}}
\end{center}
\caption{From Corsika Code: dot line is QGSJET model (Kalmykov  et al. 1993) ; dash line is VENUS model (Werner 1993) ; dash-dot line is SIBYLL model (Fletcher et al. 1994) ; short-dash line is DPMJET model ; dash-dot-dot line is HDPM model. Solid line represent our paramentrization to . 
$\sigma^{p-air}_{in}$ as in Figure 1. Solid square is from Yodh (Yodh et al. 1983). Solid circle from Mielke (Mielke et al. 1993; Mielke et al. 1994). Star from Akeno (Honda et al. 1993). Down triangle is Akeno data with Bellandi correction(Bellandi et al. 1997). Open diamond from EASTOP (Aglietta et al. 1998). Solid diamond from Fly's Eyes (Baltrusaitis et al. 1985).}
\end{figure}

\vspace{1ex}
\begin{center}
{\Large\bf References}
\end{center}
Abe, F. {\it et al.} (CDF Collab.) 1994 - Phys. Rev. {\bf D 50} 5550. \\
Aglietta, M. {\it et al.} (EASTOP Collab.) 1998\\
Amos, N.A. {\it et al.} (E710 Collab.) 1992 - Phys. Rev. Lett. {\bf 68} 2433. \\
Baltrusaitis, R.M. (Fly's Eyes Collab.) 1985 - Nucl. Instrum. and Meth. {\bf A 240}, 410. \\
Barrett, R.C. \& Jackson, D.F. 1977 - Nuclear Sizes and Structure, ( Reading: Clarendon Press, Oxford ). \\
Bellandi, J. {\it et al.} 1997 - J. of Phys. G: Nuclear and Particle Physics {\bf 23}, 125.
 \\
Caso, C. {\it et al.} (Particle Data Group) 1998 - European J. Physics {\bf 3C}. \\
Fletcher, R.S. {\it et al.} 1994 - Phys. Rev. {\bf D 50}, 5710. \\
Gaisser \\
Glauber, R.J. 1959 - Lect. Theor. Phys. Vol. 1, Edited by W. Britten and L.G. Dunhan (Reading: Interscience, NY).  \\
Glauber, R.J. {\it et al.} 1970 - Nucl. Phys. {\bf B 12}, 135. \\
Hara T. {\it et al} 1983 - Phys. Rev. Lett. {\bf 50} 2058. \\
Honda M. {\it et al} (Akeno Collab.) 1993 - Phys. Rev. Lett. {\bf 70} 525. \\
Kalmykov, N.N. \& Ostapchenki, S.S. 1993 - Phys. At. Nucl. {\bf 56}, 346. \\
Knapp, J., Heck, D., Ostapchenko, S.S., Schatz, G. 1996 (Kascade Collab. )- "Comparison of Hadronic Interaction Models used in Air Shower Simulations and of their Influence on Shower Development and Observables" - (Draft, July 31). \\
Mielke H. H. {\it et al} (Kascade Collab.) 1993 - Proc. $23^{\rm th}$ ICRC (Calgary, 1993) {\bf 4}, 155. \\
Mielke H. H. {\it et al} (Kascade Collab.) 1994 - J. Phys. G: Nucl. Part. Phys. {\bf 20} 637. \\
Werner, K. 1993 - Phys. Report {\bf 232}, 87. \\
Woods, R.D. \& Saxon, D.S. 1954 - Phys. Rev. {\bf 95}, 577. \\
Yodh, G.B. {\it et al.} 1983 - Phys. Rev. {\bf D 27}, 1183. \\

\noindent {\it Contribution to the 26$^{\rm th}$ International Cosmic Ray Conference, Salt Lake City, Utah, August 1999.}

\end{document}